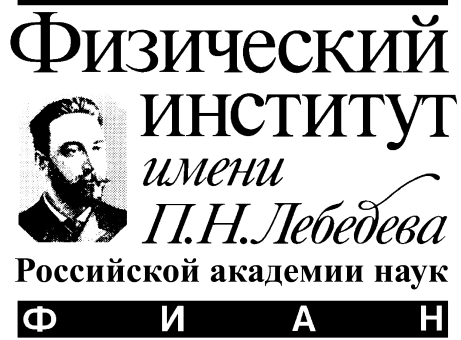

PREPRINT
**13**

A. V. BAGULYA, O. D. DALKAROV, M. A. NEGODAEV,
Yu. L. PIVOVAROV, A. S. RUSETSKII,
T. A. TUKHFATULLIN

# Orientation Effect in d(d,n)³He Reaction Initiated by 20 keV Deuterons at Channeling in Textured CVD-Diamond Target

A.V. Bagulya[1], O.D. Dalkarov[1], M.A. Negodaev[1,*], Yu.L. Pivovarov[2], A.S. Rusetskii[1], T.A. Tukhfatullin[2,**]

[1] *P.N. Lebedev Physical Institute, Russian Academy of Sciences, Moscow, Russia*
[2] *National Research Tomsk Polytechnic University, Tomsk, Russia*

**Orientation Effect in d(d,n)³He Reaction Initiated by 20 keV Deuterons at Channeling in Textured CVD – Diamond Target**

[*] Corresponding author. E-mail: negodaev@lebedev.ru

[**] Corresponding author. E-mail: tta@tpu.ru

ABSTRACT

Orientation effect of increasing the enhancement factor of DD reaction in CVD-Diamond was investigated by simulation. It is obtained that the flux peaking effect up to 2.2 times increases the relative enhancement factor for a parallel beam and up to 1.2 times for the deuteron beam with angular divergence equals 3 critical channeling angles. Qualitative agreement with the experiment was obtained.

## 1. Introduction

The interactions of the deuterium beam with deuterium enriched fixed targets are investigated in [1–4] using HELIS accelerator facility at the P.N. Lebedev Physical Institute of the Russian Academy of Sciences (LPI). In Ref. [5] the authors investigated the neutron yield in the reaction:

$$d + d \to n(2.45 \text{ MeV}) + {}^3\text{He}(0.8 \text{ MeV}) \qquad (1)$$

using textured CVD-Diamond target and 20 keV deuterium beam from HELIS accelerator which delivers the beam with small angular and energy divergences. In [5] the authors suggested that the observed enhancement of neutron yield is connected both with the screening and channeling effects.

To clarify the role of channeling in enhancement of neutron yield in d(d,n)$^3$He reaction in CVD-Diamond crystal target, we present here the results of computer simulations. The deuterons trajectories in crystal are simulated using the computer code Basic Channeling with Mathematica™ (BCM-1.0) [6], which allows calculate angular and spatial distribution of channeled particles in a thin crystals, see e.g. [7].

## 2. Experimental setup and results of experiment

Investigation of DD reaction yield was performed at HELIS accelerator facility. This multi-purpose accelerator facility operates continuous ion beams with currents up to 50 mA and energies up to 50 keV. A schematic diagram of the HELIS setup is shown in Fig. 1.

Nuclear DD reaction in the interactions of the deuterium ion beam with deuterium-enriched fixed targets were conduct using the polycrystalline deuterium-enriched of 400 μm thick polycrystalline CVD diamond. The method of target formation was the follows. The film was grown on a 57 mm diameter silicon substrate using the microwave plasma-assisted chemical vapour deposition (MPCVD) system [8].

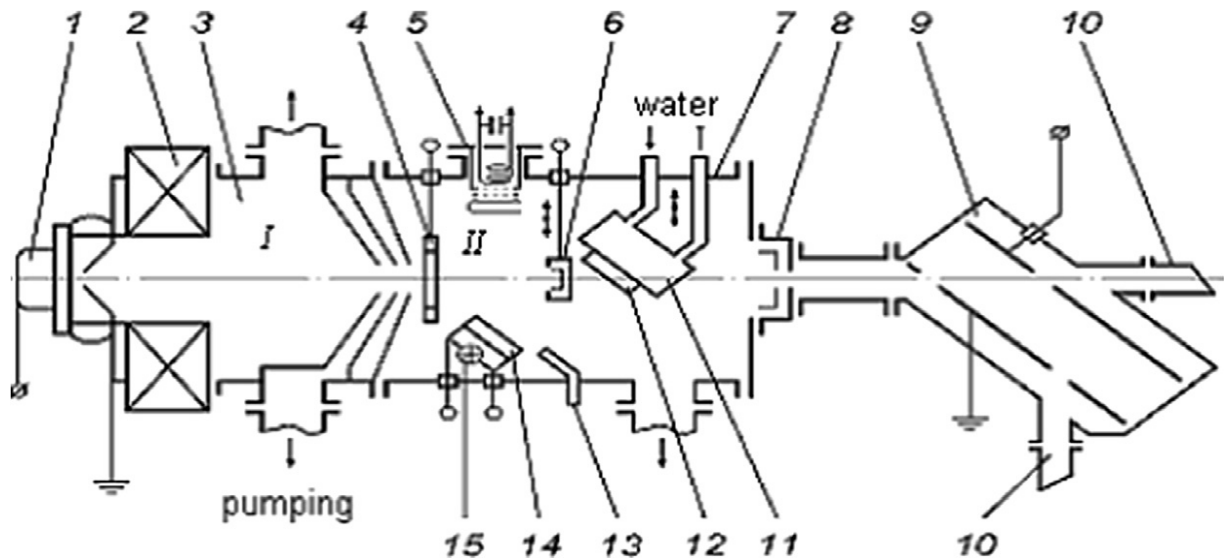

**Fig. 1**. Schematic diagram of the HELIS setup:
1 – the ion source (duoplasmatron); 2 – electromagnetic lens; 3 – three-stage chamber of differential pumping; 4 – non contact current meter; 5 – auxiliary ion source; 6 and 10 – Faraday cages; 7 – targets chamber; 8 – the device for calorimetric measurement of the ion beam current; 9 – electrostatic analyzer; 11 – water cooled target holder; 12 – target; 13 – feeder of gas in vacuum chamber; 14 – substrate; 15 – substrate heater.

The black diamond film was obtained, with numerous structural defects in the crystallites, such as twins and amorphous carbon inclusions with a size up to 1 nm. Further the diamond film was separated from the substrate by etching of silicon in a mixture of hydrofluoric nitric and acetic acids, and cut by a Nd: YAG laser into discs of 18 mm diameter. The structure of polycrystalline diamond is anisotropic and not homogeneous. The crystallites are growing in the form of columns, perpendicular towards the surface. The transversal size of crystallites increases from ≈1 μm in the layer close to the substrate to about 50 μm on the growth side. The growth surface of the sample shows a clear crystalline structure with (100) grain orientation.

The neutron flux, produced in the DD reaction, was measured in the longitudinal and transverse direction with respect to the beam axis by using a multichannel neutron detector based of $^3$He counters. The relative yield of the DD reaction was determined as $Y_{dd} = n_n/(S\ I_d)$, where $n_n$ is the neutron flux, $S$ is the irradiated area of the target and $I_d$ denotes the ion beam current. The observed neutron yield measured in longitudinal and transverse directions with respect to the ion beam as a function of the angle between the beam direction and the norm to the target plane is shown in Fig. 2. The enhancement of the DD reaction yield with decreasing angle is clearly observed.

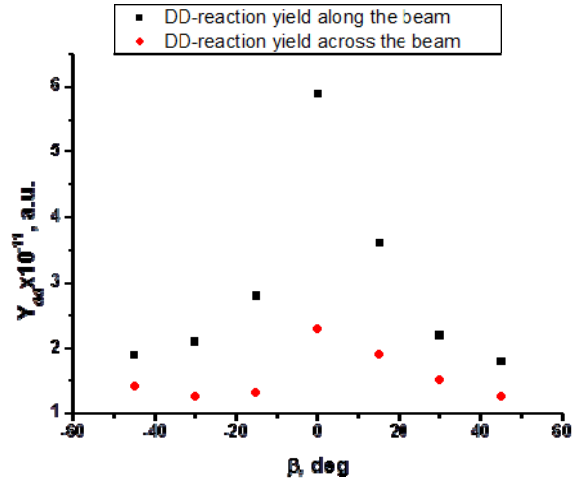

**Fig. 2**. The relative neutron yield obtained with the CVD-diamond sample as a function of the angle β between the beam and the target plane norm, measured in longitudinal (■) and transverse (♦) directions with respect to the ion beam. Ion beam with the energy of $E_d$=20 keV and the current of 50 μA [5].

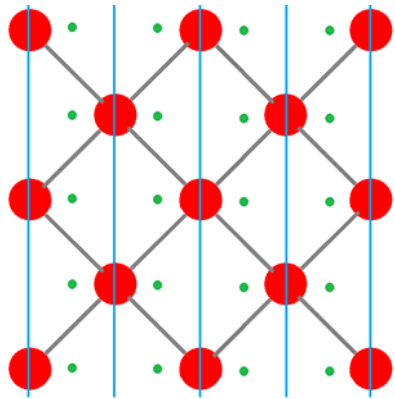

**Fig. 3.** Diamond lattice for <100> direction. Small green circles are positions of deuterium atoms, and large red circles are positions of carbon atoms. Blue vertical line are the (400) channeling planes.

## 3. Simulations results and discussion

In the first calculation we suggest that deuterium atoms are arranged in perfect diamond lattice between the (400) planes (see in Fig. 3.). In this case, channeled deuterium ions move mainly between the channeling plane and passes near the deuterium atoms situated in the target, which should lead to the enhancement of the DD reaction.

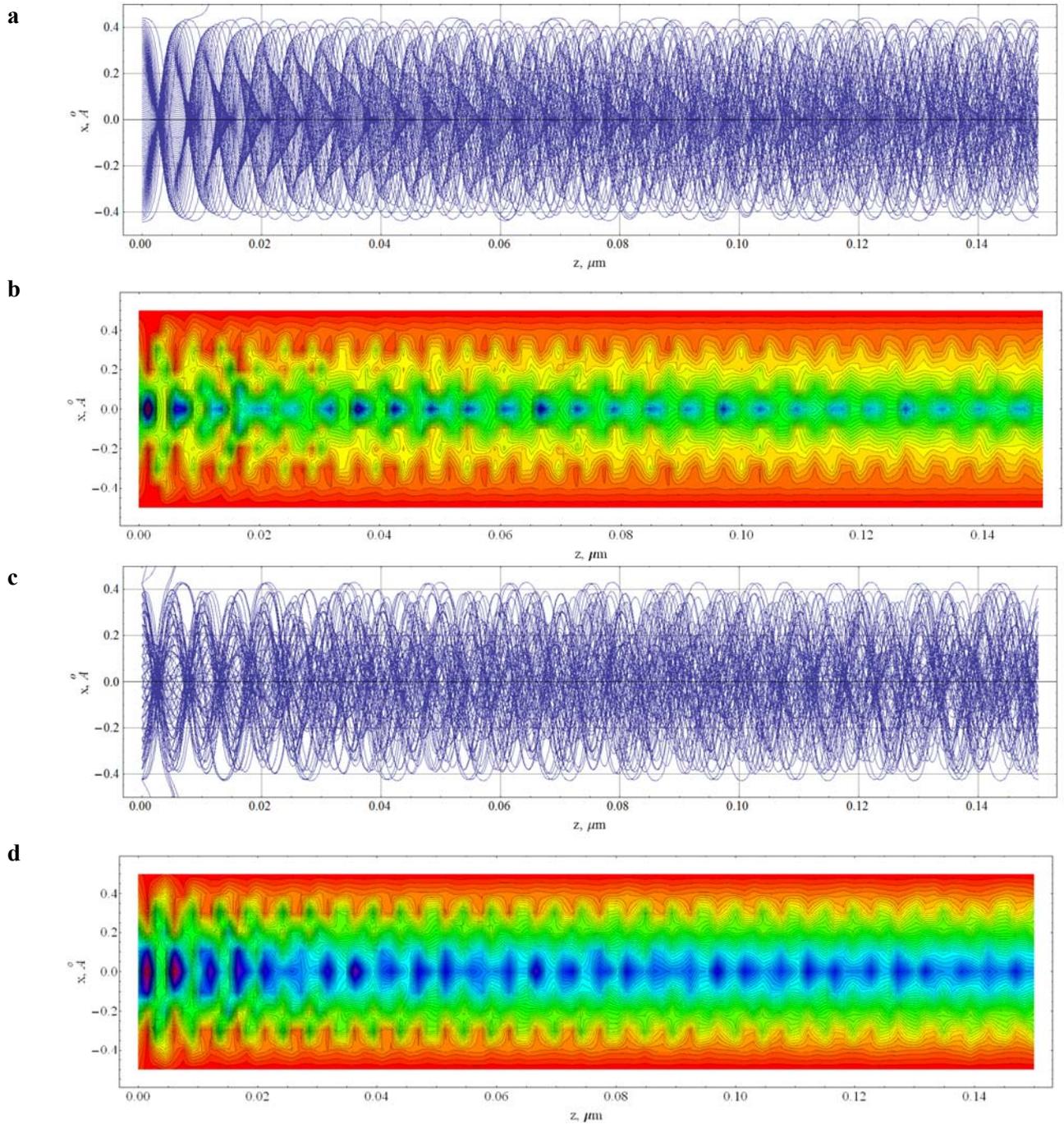

**Fig. 4.** Simulated trajectories of 20 keV deuterium ions in (400) C crystal and corresponding flux density:
(a), (b) parallel beam,
(c), (d) beam with angular spread $\Delta\theta=\theta_c/5$ ($\theta_c=1.4^o$ is the critical channeling angle).

Simulations of the ions trajectories are carried out using the code "Basic Channeling with Mathematica" BCM–1.0 [6]. This code computes numerical solutions of the classical equations of motion using the model of continuum potential. Results of calculation of trajectories for 20 keV deuterium ions in a C crystal channeled along the (400) plane are depicted in Fig. 4. Because the potential for planar-channeled positive particles is close to harmonic, the trajectories are characterized by a single oscillation wavelength $\lambda$ and the oscillation period weakly depend on initial transversal coordinate. So the bright flux peaking effect is clearly observed. For the crystal thickness equals to 0.15 μm we observe the 10 maxima on the flux density function for ideal parallel beam (Fig. 4b). Due to these maxima we expect the enhancement of the DD reaction yield.

Computer code allows simulating the beam with difference angular spread; thereby we take into account the angular divergence of the real ion beam and the fact that in textured CVD-Diamond the crystallites can be oriented differently with respect to the target surface as well. Thus in the simulation for each entry points of deuterium ion into the crystal the five incidence angle was generated using random real generator with the normal distribution and specific standard deviation $\Delta\theta$. Results of simulation of trajectories for 20 keV deuterium ions with angular spread $\Delta\theta = \theta_c/5$ ($\theta_c=1.4^o$ is the critical channelling angle) are shown in the Fig. 4c, d. Because of the the oscillation period weakly depends on the initial transvers energy of deuterons the flux maxima on spatial distribution are not sharp as for the case of parallel beam, that should lead to the decrease of the enhancement factor (EF) of reaction.

Probability of the DD reaction depends on the impact parameter of the deuterium ions, penetrating trough the crystal with respect to the deuterium atoms situated in the target. To estimate the relative EF of DD reaction it is necessary to estimate the number of trajectories passing close to the center of the channel where target deuterium atoms are located:

$$R(\theta) = \frac{\int_0^L dz \sum_{-n}^{n} \int_{-\Delta x}^{\Delta x} f_{ch}(x + nd_p, z) dx}{\int_0^L dz \sum_{-n}^{n} \int_{-\Delta x}^{\Delta x} f_{unch}(x + nd_p, z) dx}, \qquad (2)$$

where $f_{ch}(x,z)$ is the flux density of channeled ions in C crystal, $f_{unch}(x,z)$ is the flux density of ions moving in amorphous target, $L$ is the crystal thickness, $n$ is the number of channel, $d_p$ is the interplanar distance. $\Delta x = 0.1$ Å is the interval inside which we count the number of trajectories.

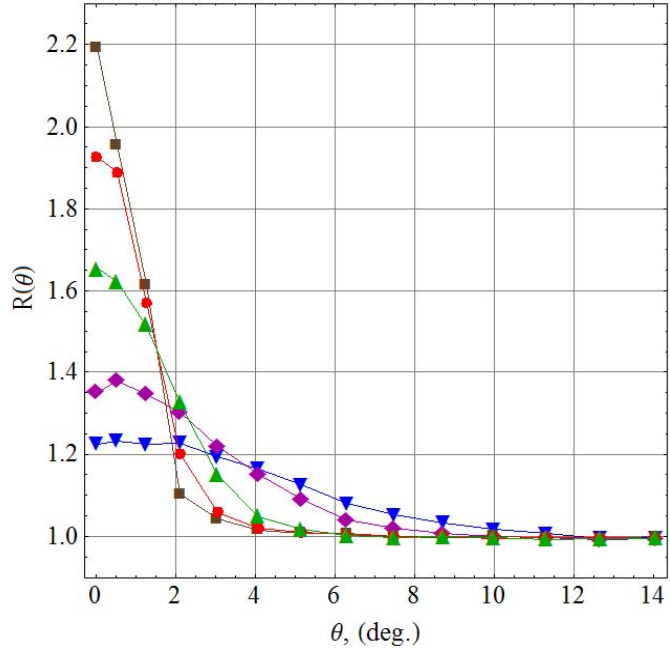

**Fig. 5**. Relative EF of the DD reaction versus angle of incidence D$^+$ ions into the C crystal for different angular spreed of the beam:
■ $\Delta\theta = \theta_c/5$, ● $\Delta\theta = \theta_c/2$, ▲ $\Delta\theta = \theta_c$, ◆ $\Delta\theta = 2\theta_c$, ▼ $\Delta\theta = 3\theta_c$
($\theta_c = 1.39°$ is the critical channeling angle).

A charged particle traversing an amorphous target is deflected by many small-angle scattering events. Most of this deflection is due to Coulomb scattering from nuclei and therefore is called multiple Coulomb scattering (MS). The MS distribution is Gaussian for small deflection angles. Thus $f_{unch}(x,z)$ is defined as a Gaussian function with standard deviation $\Delta\theta_{unch} = \sqrt{\Delta\theta_{MS}^2 + \Delta\theta^2}$, where $\Delta\theta$ is angular spread of the ion beam and $\Delta\theta_{MS}$ is the angular divergence due to MS. According [9] for diamond crystal and deuterium ions with energy $E$=20 keV $\Delta\theta_{MS}$=10.4°.

The calculation result of relative EF of DD reaction as function of incidence angle of $D^+$ ion with respect to channeling plane is shown on the Fig. 5. The relative EF increased with decreasing incident angle of deuterium ions with respect to channeling plane because in this case it is increasing the number of ions involved in channeling motion. The similar dependence of neutron yield was observed in the experiment [5]. Flux peaking effect up to 2.2 times increases the EF of DD reaction for beam with angular divergence $\Delta\theta = \theta_c/5$, and it is reduces to 1.2 with increasing angular divergence of ion beam to $\Delta\theta = 3\theta_c$.

## 4. Conclusions

Orientation effect of increasing the EF of DD reaction in CVD-Diamond was investigated by simulation. It is obtained that the flux peaking effect up to 2.2 times increases the relative EF for a parallel beam and up to 1.2 times for the deuteron beam with angular divergence equals 3 critical channeling angles. Qualitative agreement with the experiment [5] was obtained.

**Acknowledgements**

The theoretical work was supported by National Research Tomsk Polytechnic University grant No VIU-NRiI-23/2016.

Александр Васильевич БАГУЛЯ
Олег Дмитриевич ДАЛЬКАРОВ
Михаил Александрович НЕГОДАЕВ
Юрий Леонидович ПИВОВАРОВ
Алексей Сергеевич РУСЕЦКИЙ
Тимур Ахатович ТУХВАТУЛЛИН

**Ориентационный эффект в реакции d(d,n)³He вызванный каналированнием дейтронов с энергией 20 кэВ в текстурированной мишени из CVD алмаза**